\documentclass[twocolumn,showpacs,aps,prd,amsmath,amssymb,nofootinbib,nobibnotes,floatfix]{revtex4-1}
\usepackage{hyperref}
\usepackage{amsfonts}
\usepackage{amsmath}
\usepackage{mathrsfs}
\usepackage{epsfig}
\usepackage{graphicx}
\usepackage{url}
\usepackage{hyperref}
\usepackage{float}
\usepackage{color}
\usepackage[utf8]{inputenc} 
\usepackage{graphicx}
\usepackage{epsf}
\usepackage{comment}
\usepackage{slashed}
\usepackage{footmisc}
\usepackage{setspace}

\begin{document}

\title{Eccentric Black Hole Mergers Forming in Globular Clusters}

\author{Johan Samsing}
\email{Email: jsamsing@gmail.com}
\affiliation{Department of Astrophysical Sciences, Princeton University, Peyton Hall, 4 Ivy Lane, Princeton, NJ 08544, USA.}

\begin{abstract}

We derive the probability for a newly formed binary black hole (BBH) to undergo an eccentric gravitational wave (GW) merger during binary-single
interactions inside a stellar cluster. By integrating over the hardening interactions such a BBH must undergo before ejection,
we find that the observable rate of BBH mergers with eccentricity $>0.1$ at $10\ \rm{Hz}$ relative to the rate of circular mergers can be as high as $\sim 5\%$ for a
typical globular cluster (GC). This further suggests that BBH mergers forming through GW captures in binary-single interactions, eccentric or not,
are likely to constitute $\sim 10\%$ of the total BBH merger rate from GCs. Such GW capture mergers can only be probed with an $N$-body code that
includes General Relativistic corrections, which explains why recent Newtonian cluster studies not have been able to resolve this population.
Finally, we show that the relative rate of eccentric BBH mergers depends on the compactness of their host cluster,
suggesting that an observed eccentricity distribution can be used to probe the origin of BBH mergers.

\end{abstract}

\maketitle

\section{Introduction}

Gravitational waves (GWs) from merging binary black holes (BBHs) have been observed \citep{2016PhRvL.116f1102A,
2016PhRvL.116x1103A, 2016PhRvX...6d1015A, 2017PhRvL.118v1101A, 2017PhRvL.119n1101A},
but their astrophysical origin is still unknown. Several formation channels and sites have been proposed in the literature, including
stellar clusters \citep{2000ApJ...528L..17P, 2010MNRAS.402..371B, 2013MNRAS.435.1358T, 2014MNRAS.440.2714B,
2015PhRvL.115e1101R, 2016PhRvD..93h4029R, 2016ApJ...824L...8R, 2016ApJ...824L...8R, 2017MNRAS.464L..36A, 2017MNRAS.469.4665P}, isolated
field binaries \citep{2012ApJ...759...52D, 2013ApJ...779...72D, 2015ApJ...806..263D, 2016ApJ...819..108B, 2016Natur.534..512B}, galactic
nuclei \citep{2009MNRAS.395.2127O, 2015MNRAS.448..754H, 2016ApJ...828...77V, 2016ApJ...831..187A, 2017arXiv170609896H}, active galactic nuclei
discs \citep{2017ApJ...835..165B,  2017MNRAS.464..946S, 2017arXiv170207818M}, as well as primordial
black holes \citep{2016PhRvL.116t1301B, 2016PhRvD..94h4013C, 2016PhRvL.117f1101S, 2016PhRvD..94h3504C}, however, how
to observationally distinguish them from each other has shown to be a major challenge. 
For this, several recent studies have explored to which degree the distributions of BBH spins and orbital eccentricities might differ
between different models \citep{2016ApJ...832L...2R, 2017ApJ...842L...2C},
as these are quantities that can be extracted from the observed GW waveform \citep{2016PhRvD..94b4012H, 2017PhRvD..95b4038H, 2017arXiv170510781G, 2018PhRvD..97b4031H}.
In general, for BBH merges evolved in isolation one finds the spins to be preferentially aligned with the
orbit \citep{2000ApJ...541..319K} and eccentricity to be indistinguishable from zero, whereas dynamically assembled BBH mergers
will have random spin orientations, and a non-zero probability for appearing
eccentric at observation \citep{2006ApJ...640..156G, 2009MNRAS.395.2127O, 2014ApJ...784...71S, 2016PhRvD..94h4013C}.
For such studies it has especially become clear that implementing General Relativistic (GR) effects is extremely important, .e.g,
GR precession and spin-orbit coupling affect both the eccentricity \citep{2013ApJ...773..187N} and the
BBH spins \citep{2017ApJ...846L..11L} in secular evolving systems, where GW emission in few-body scatterings is essential for
resolving the fraction of highly eccentric mergers \citep{2006ApJ...640..156G, 2017ApJ...840L..14S}. Despite this importance,
many recent studies are still based on purely Newtonian codes.

In this paper we study the evolution of BBHs undergoing hardening binary-single interactions inside a dense stellar cluster, 
and how the inclusion of GR corrections affect both the dynamical history of the BBHs and their GW merger distribution.
We especially follow the GW mergers that form {\it during} the hardening binary-single
interactions through GW captures \citep[\textit{e.g.},][]{2006ApJ...640..156G, 2014ApJ...784...71S}.
By integrating over the binary-single interactions a typical BBH undergoes inside its host cluster,
we derive that the rate of BBH mergers forming during binary-single interactions with an eccentricity $>0.1$ at $10\ \rm{Hz}$ (eccentric mergers)
relative to the rate of classically ejected BBH mergers (circular mergers) can be as high as $\sim 5\%$ for a typical GC. This rate is within observable
limits, suggesting that the eccentricity distribution of BBH mergers can be used to constrain their origin. We note that the binary-single GW captures that lead to this large
fraction of eccentric mergers only can be probed when GR effects are included in the $N$-body equation-of-motion (EOM), which
explains why recent Newtonian Monte-Carlo (MC) cluster studies not have been able to resolve this population
\citep[\textit{e.g.},][]{2016PhRvD..93h4029R, 2016ApJ...816...65A}. In fact, we explicitly prove in this paper that a Newtonian code will always
underestimate the eccentric fraction by a factor of $\sim 100$.

Our present study further suggests that GW capture mergers forming during three-body interactions,  eccentric or not,
are likely to constitute $\sim 10 \%$ of the total observable BBH merger rate from GCs. This population is currently unexplored, but is likely to play
a key role in constraining the time dependent dynamical state of BHs in clusters, as it might leave unique imprints across frequencies observable
by both the `Laser Interferometer Space Antenna' (LISA) and the `Laser Interferometer Gravitational-Wave Observatory' (LIGO).

Throughout the paper we assume that all three interacting BHs have the same mass $m$, and that the total initial energy of the
three-body system is dominated by that of the initial target binary; a limit formally known as the hard binary (HB) limit \citep{Heggie:1975uy, Hut:1983js}.
We only discuss effects from dynamical GW emission, which appears in the post-Newtonian (PN) expansion formalism at the 2.5 order \citep{2014LRR....17....2B}.
The lower PN terms leading to precession are important for describing secular systems \citep{2016ARA&A..54..441N}, but not the chaotic ones we
consider in this work \citep{2006ApJ...640..156G, 2014ApJ...784...71S}.

\section{Eccentric Capture Distances}\label{sec:Eccentric Capture Distances}

There are two characteristic pericenter distances related to the formation of eccentric BBH mergers: the distance at which the
GW peak frequency of a BBH has a certain value $f$, denoted by $r_{f}$, and the distance from which a BBH can undergo a GW capture and
still have a non-negligible eccentricity $e_{f}$ when its GW peak frequency is $f$, denoted by $r_{\rm EM}$, where `EM' is short for `Eccentric Merger'.
In the resonating three-body problem \citep{2014ApJ...784...71S}, a third relevant distance also exists, namely the characteristic distance from which two of the
three interacting BHs will be able to undergo a GW capture
during the interaction without being interrupted by the bound single, referred to as $r_{\rm cap}$. As shown in
\citep{2018ApJ...853..140S}, the distance $r_{\rm cap}$ does not equal a constant value,
in contrast to $r_{f}$ and $r_{\rm EM}$, but differs between each of the temporarily lived BH pairs, also referred to as
intermediate state (IMS) BBHs \citep{2014ApJ...784...71S, 2018ApJ...853..140S}, assembled during the resonating three-body state. In this paper we assume that
$r_{\rm cap} > r_{\rm EM}$, i.e., we work in the limit where all IMS BBHs with pericenter distance $r_{\rm p} \leq r_{\rm EM}$ also
undergo a GW capture merger. This is an excellent approximation for LIGO sources, but not necessarily for LISA sources, due to their difference
in frequency sensitivity.
In the following three paragraphs we estimate $r_{f}$ (GW frequency distance), $r_{\rm EM}$ (Eccentric merger distance),
and $r_{\rm cap}$ (GW capture distance), respectively. For further descriptions of the resonating three-body problem with and without GR
we refer the reader to \citep{2014ApJ...784...71S, 2017ApJ...840L..14S, 2017ApJ...846...36S, 2018ApJ...853..140S, 2017arXiv170604672S, 2017arXiv170901660S}.

\begin{figure}
\centering
\includegraphics[width=\columnwidth]{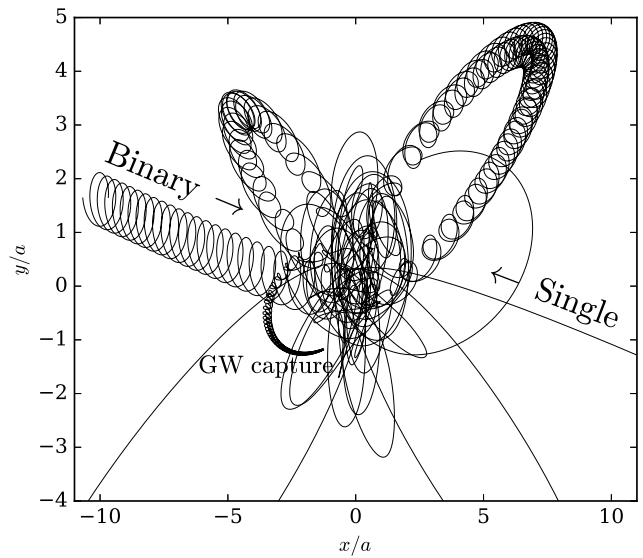}
\caption{Formation of an eccentric BBH GW merger during a resonating binary-single interaction between three equal mass BHs. The location of
the eccentric GW capture merger is denoted by `GW capture', where the initial paths of the incoming BBH and single BH, are denoted
by `Binary' and `Single', respectively. The GW capture forms as a result of GW emission during a close encounter between two of the three BHs
while they temporarily form a bound three-body state. Such GW capture mergers often appear highly eccentric at $10\ \rm{Hz}$.
}
\label{fig:resGWcapex}
\end{figure}

\subsection{GW frequency distance $r_{f}$}

The GW peak frequency $f$ of a BBH with SMA $a$ and eccentricity $e$, can be approximated by that found from assuming the two BHs
are on a circular orbit with a SMA equal to the pericenter distance $r_{\rm p} = a(1-e)$ \citep{Wen:2003bu}. Using that the emitted GW frequency is
two times the Keplerian orbital frequency follows directly that $f \approx \pi^{-1} \sqrt{{2Gm}/{r_{f}^3}}$.
For a BBH to emit GWs with peak frequency $f$, its pericenter distance must therefore be,
\begin{equation}
{r_{f}} \approx \left( \frac{2Gm}{f^2 {\pi}^2}\right)^{1/3}.
\label{eq:rf}
\end{equation}
As a result, if a BBH has a pericenter distance $r_{\rm p} \leq r_{f}(f = 10\ \text{Hz})$ then it will emit GWs at a frequency $f \geq 10\ \text{Hz}$
and therefore be immediately observable by an instrument similar to LIGO. As the relevant distance $r_{f}$ for LIGO is $\ll a$ for all realistic
astrophysical systems, the corresponding BBH eccentricity will therefore be extremely high, as indeed found using numerical PN scattering
experiments \citep{2017ApJ...840L..14S}. Such GW sources are said to be born in the LIGO band \citep{2017ApJ...840L..14S}.

\subsection{Eccentric merger distance $r_{\rm EM}$}

A BBH that forms with an initial pericenter distance $r_{\rm p} >  r_{f}$ is not immediately observable at GW frequency $f$.
For that, its pericenter distance must decrease, which naturally happens through GW emission during inspiral \citep{Peters:1964bc}.
However, in that process, the BBH also undergoes significant circularization \citep{Peters:1964bc}, and will
as a result generally appear with a relative low eccentricity once the GW peak frequency is $f$.
To estimate the characteristic pericenter distance $r_{\rm EM}$ for which the eccentricity is $e_{f}$ at frequency $f$, we make use of the analytical
relation between the time evolving pericenter distance and eccentricity derived in \cite{Peters:1964bc},
\begin{equation}
r_{\rm p}(e) = r_{f} \times {F(e)}/{F(e_{f})},
\label{eq:rp_e}
\end{equation}
where $F(e)$ denotes the function,
\begin{equation}
F(e) = \frac{e^{12/19}}{1+e} \left(1 + \frac{121}{304}e^2 \right)^{870/2299}.
\end{equation}
We have here normalized the expression for $r_{\rm p}(e)$ such that $r_{\rm p} = r_{f}$ when $e = e_{f}$. Using that the eccentricity of a typical
IMS BBH at the time of its formation is close to unity, as $r_{\rm EM} \ll a$, one finds that $r_{\rm EM}$ is simply given by
Equation \eqref{eq:rp_e} evaluated in the limit for which $e \rightarrow 1$,
\begin{equation}
r_{\rm EM} \approx r_{f} \times \frac{1}{2F(e_{f})} \left(\frac{425}{304}\right)^{870/2299}.
\label{eq:r_EM}
\end{equation}
For $e_{f} = 0.1$ follows that $r_{\rm EM}/r_{f} \approx 2.7$, i.e., GW capture mergers with an initial $r_{\rm p}$ up to about three times the distance $r_{f}$
will appear eccentric at the time of observation for  an instrument similar to LIGO. Note here that
this ratio is independent of the frequency $f$.

\subsection{GW capture distance $r_{\rm cap}$}\label{sec:GW capture distance}

The characteristic pericenter distance from which two of the three interacting BHs can undergo a GW capture merger, $r_{\rm cap}$, is that
for which the GW energy loss integrated over one pericenter passage, $\Delta{E}_{\rm p}(r_{\rm p}) \approx (85\pi/12)G^{7/2}c^{-5}m^{9/2}r_{\rm p}^{-7/2}$ (see \cite{Hansen:1972il}), is
comparable to the total energy of the three-body system \citep{2017ApJ...846...36S, 2018ApJ...853..140S} that in the HB limit is that
of the initial target binary, $E_{\rm B}(a) \approx Gm^{2}/(2a)$ (see \cite{2014ApJ...784...71S}).
Solving for the pericenter distance for which $\Delta{E}_{\rm p}(r_{\rm cap}) = E_{\rm B}(a)$, one now finds \citep{2017ApJ...846...36S},
\begin{equation}
r_{\rm cap} \approx {\mathscr{R}_{\rm m}} \times \left({a}/{{\mathscr{R}_{\rm m}}}\right)^{2/7},
\label{eq:rcap}
\end{equation}
where ${\mathscr{R}_{\rm m}}$ denotes the Schwarzschild radius of a BH with mass $m$. As described in the introduction to Section \ref{sec:Eccentric Capture Distances},
$r_{\rm cap}$ is not a fixed distance, but varies throughout the resonating state \citep{2014ApJ...784...71S, 2018ApJ...853..140S}, the normalization
of the estimate given by the above Equation \eqref{eq:rcap} is therefore only approximate. However, to get a sense of the relevant scale, one finds for $m=20M_{\odot}$ and $a = 1\text{\ au}$
that $r_{\rm cap}/{\mathscr{R}_{\rm m}} \approx 100$, i.e., for these values if two of the three BHs pass each other within a distance
of $\sim 100 \times {\mathscr{R}_{\rm m}}$, then they are likely to undergo a GW capture merger. For a more extensive solution and description of the problem,
where the varying capture distance is taken into account, we refer the reader to \citep{2018ApJ...853..140S}.
An example of a GW capture forming during a resonating binary-single interaction is shown in Figure \ref{fig:resGWcapex}.

\section{Eccentric Merger Probability}\label{sec:Eccentric Merger Probability}

The total probability for a BBH to undergo an eccentric GW capture merger during binary-single
interactions (Figure \ref{fig:resGWcapex}) inside a cluster (Figure \ref{fig:BS_cluster_ill}), can be estimated by
simply summing up the probability for each of the hardening interactions the BBH must undergo before 
ejection from the cluster is possible. In the sections below we estimate this integrated probability, show how it depends on the properties of the host cluster,
and compare it to other BBH merger types.
For our calculations we assume that the probability for the BBH in question to undergo a merger before ejection is possible $\ll 1$,
which allow us to express the total probability for any merger type as a simple uncorrelated sum over the interactions.
As later derived in Section \ref{sec:Isolated Merger Before Ejection}, and illustrated in
Section \ref{sec:Rate of Eccentric Mergers}, this assumption is valid for standard GC systems, but will break down for dense nuclear star clusters.
The process of BBH hardening and cluster ejection is further illustrated and described in Figure \ref{fig:BS_cluster_ill}.

\subsection{A single interaction}

We first estimate the probability for an IMS BBH to form and undergo a GW capture merger
with an initial $r_{\rm p} \leq r_{\rm EM}$, during an interaction between a BBH with initial SMA $a$, and a single incoming BH. We generally
refer to this probability as $P_{\rm EM}(a)$.
For this, we start by noting that the SMA of each formed IMS BBH, denoted by $a_{\rm IMS}$, is similar to the SMA of the initial target
binary, i.e., $a_{\rm IMS} \approx a$.
For a BBH to form with an initial $r_{\rm p} < r_{\rm EM}$ its eccentricity at formation must therefore be $> e_{\rm EM}$, where
$e_{\rm EM} = 1 - r_{\rm EM}/a$. The probability for a single IMS BBH to form with $r_{\rm p} < r_{\rm EM}$ is therefore
equal to that of forming with $e > e_{\rm EM}$, which is given by $(1-e_{\rm EM}^2) \approx 2(1-e_{\rm EM}) = 2 r_{\rm EM}/a$,
under the assumption that the eccentricity distribution follows a so-called thermal distribution $P(e) = 2e$ \citep{Heggie:1975uy}. By weighting with the average
number of IMS BBHs forming during a HB binary-single interaction, denoted here by $N_{\rm IMS}$, one now finds,
\begin{equation}
P_{\rm EM}(a) \approx \frac{2r_{\rm EM}}{a} \times N_{\rm IMS}.
\label{eq:Pef_a}
\end{equation}
We note here that $N_{\rm IMS}$ in the collisionless non-relativistic HB limit is independent of both the absolute mass scale
and the initial SMA \citep{1983AJ.....88.1549H, Hut:1983js}.
As $r_{\rm cap} \ll a$, we can therefore take $N_{\rm IMS}$ to be constant in this work. Its value
can be analytically estimated by using that the normalized orbital energy distribution of binaries assembled in three-body
interactions approximately follows \citep{Heggie:1975uy, 2006tbp..book.....V},
\begin{equation}
P(E_{\rm B}) \approx ({7}/{2})E_{\rm B}(a)^{7/2} \times E_{\rm B}^{-9/2},
\label{eq:PEB}
\end{equation}
Following this approach, the number $N_{\rm IMS}$ is simply equal to the probability for an assembled BBH to have
$E_{\rm B} < E_{\rm B}(a)$ (single is bound) divided by the probability for $E_{\rm B} > E_{\rm B}(a)$ (single is unbound).
These probabilities can be found from integrations of Equation \eqref{eq:PEB}, from which
follows that $N_{\rm IMS} \approx (\max(a_{\rm IMS})/a)^{7/2}$, where $\max(a_{\rm IMS})$ denotes the maximum value of $a_{\rm IMS}$.
The ratio $\max(a_{\rm IMS})/a$ is between $2-3$ (an exact value cannot be derived, as our framework breaks down when
the three-body state no longer can be described by a binary with a bound single \citep{2018ApJ...853..140S}), which then
translates to an $N_{\rm IMS}$ between $10 \sim 40$.
Using a large set of isotropic three-body scatterings we determined its average value to be $N_{\rm IMS} \approx 20$, which
is the value we will use throughout the paper.

\subsection{Integrating over hardening interactions}

The majority of BBHs in a cluster are formed with an initial $a$, denoted by $a_{\rm in}$, that is greater than the
maximum $a$ that leads to a dynamical ejection of the BBH out of the cluster through a binary-single interaction (we determine this value
later in the paper), a value we refer to as $a_{\rm ej}$.
A newly formed BBH will therefore typically have to undergo several hardening binary-single interactions, each of which slightly decreases its SMA, before
ejection from the cluster is possible. During each of these interactions there is a finite probability for two of the three
BHs to undergo an eccentric GW capture merger, implying that the relative number of eccentric mergers forming per
BBH is larger than the number evaluated at, e.g., $a_{\rm ej}$. The eccentric merger fraction must therefore be larger than the recently
reported $1 \sim 2\%$ by \citep{2017ApJ...840L..14S, 2018ApJ...853..140S}. In the paragraphs below we estimate the expected
increase from including the dynamical hardening process.

\subsubsection{Binary-single hardening process}\label{sec:Binary-single hardening process}

\begin{figure}
\centering
\includegraphics[width=\columnwidth]{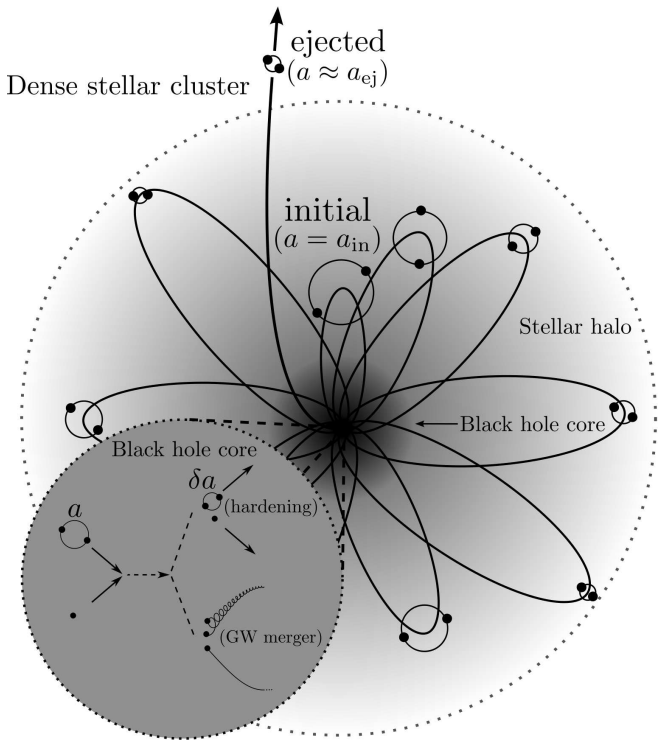}
\caption{Illustration of a BBH undergoing hardening binary-single interactions in a stellar cluster.
Initially the BBH (labeled by `initial') forms with a SMA $a_{\rm ej} < a_{\rm in} < a_{\rm HB}$, either dynamically or primordially, after which it
sinks to the core due to dynamical friction. The BBH here undergoes a HB binary-single interaction, which classically concludes with the
BBH receiving a kick velocity $v_{\rm B}$ that unbinds it from the single and sends it back into the cluster. It then sinks back to the core, after which the
process repeats. Each of these HB binary-single interactions gradually decreases the SMA of the BBH,
which correspondingly leads to increasing dynamical kicks. When the SMA of the BBH reaches $a \approx a_{\rm ej}$, i.e., when the
dynamical kick velocity is about the escape
velocity of the cluster, then the following binary-single interaction will eject the BBH out of the cluster (labeled by `ejected'), after which it merges
in isolation.
However, if GW emission is included in the $N$-body solver, then the BBH can also undergo a GW capture merger inside the cluster core
during one of its hardening binary-single interactions, as illustrated in Figure \ref{fig:resGWcapex}. 
The grey insert circle shows a zoom in on the core region. As described, the BBH here undergoes binary-single interactions that either will lead to
hardening (the SMA changes from $a$ to $\delta a$, labeled `hardening'), or a GW merger during the interaction if GR effects are included (labeled `GW merger').}
\label{fig:BS_cluster_ill}
\end{figure}

We start by considering a single BBH, and assume that its SMA per interaction changes
from $a$ (before the binary-single interaction) to $\delta a$ (after the interaction), where $\delta < 1$ (see Figure \ref{fig:BS_cluster_ill}).
We note here that $\delta$ can be considered a constant in the HB limit, due to the scale free nature of the problem \citep{Hut:1983js}. 
A representative value for $\delta$ can be found by the use of the binary energy distribution $P(E_{\rm B}) \propto E_{\rm B}^{-9/2}$
introduced in Equation \eqref{eq:PEB}. By changing variable from $E_{\rm B}$ to $\delta$, one finds that the mean
value of $\delta$, denoted here by $\langle \delta \rangle$, is given by,
\begin{equation}
\langle \delta \rangle = \frac{7}{2} \int_{0}^{1} {\delta}^{7/2} d\delta = \frac{7}{9}.
\label{eq:ave_delta}
\end{equation}
For simplicity, we will therefore use $\delta = 7/9$ throughout the paper when evaluating actual numbers; however, we do note
that to estimate the true expectation values of the different observables we consider in this paper the full distribution of $\delta$ must
in principle be used. This is not easy, but we do hope to improve on this in upcoming studies. Finally, it is worth noting that the average
value of $E_{\rm B}$, found by simply integrating over $E_{\rm B}P(E_{\rm B})$,
is given by $\langle E_{\rm B} \rangle = (7/5)E_{\rm B}(a)$, which implies that the average fractional increase in binding energy per
binary-single interaction is $7/5-1 = 0.4$. This estimate is in full agreement with that found from numerical scatterings experiments
\citep{2017MNRAS.469.4665P}, which validates at least this part of our approach.

Following this approach, each binary-single interaction therefore releases an amount of energy equal
to $\Delta{E_{\rm bs}} = (1/\delta - 1) \times E_{\rm B}(a)$, which relates to the
recoil velocity the BBH receives in the three-body center-of-mass (COM) as $\Delta{E}_{\rm bs} = 3mv_{\rm B}^2$, where $v_{\rm B}$ is the BBH recoil velocity
defined at infinity. When $a$ is such that $v_{\rm B} > v_{\rm esc}$, where $v_{\rm esc}$ denotes the escape velocity of the
cluster, then, per definition, the BBH escapes. By assuming that $v_{\rm esc}$ is about the velocity dispersion of the cluster, one can write the ratio between the
HB limit for $a$ \citep{Hut:1983js}, denoted by $a_{\rm HB}$, and the ejection value $a_{\rm ej}$ by,
\begin{equation}
\frac{a_{\rm HB}}{a_{\rm ej}} \approx \frac{9}{1/\delta - 1}.
\label{eq:aHB_aej}
\end{equation}
We note here that this is a lower limit as $v_{\rm esc}$ in general is slightly greater than the dispersion value.
For $\delta = 7/9$ one finds ${a_{\rm HB}}/{a_{\rm ej}} \approx 30$, i.e.,
a binary formed with $a = a_{\rm HB}$ needs to decrease its SMA by a factor of $\sim 30$ before its binding energy is large enough for the
three-body recoil to eject it form the cluster.

Finally, the number of binary-single interactions required to bring a BBH from $a_{\rm in}$ to $a_{\rm ej}$, denoted by
$N_{\rm bs}(a_{\rm in}, a_{\rm ej})$, is given by,
\begin{equation}
{N_{\rm bs}}(a_{\rm in}, a_{\rm ej}) = \int_{a_{\rm ej}}^{a_{\rm in}} \frac{1}{1-\delta} \frac{1}{a} da = \frac{1}{1-\delta} \ln \left(\frac{a_{\rm in}}{a_{\rm ej}}\right),
\end{equation}
where we have used that $da = -a(1-\delta)dN_{\rm bs}$. For $\delta = 7/9$ one finds that
${N_{\rm bs}}(a_{\rm HB}, a_{\rm ej}) \approx 15$, which illustrates the point that a BBH formed in a cluster generally
undergoes a non-negligible number of scatterings before ejection (see \citep{1992ApJ...389..527H, 2017MNRAS.469.4665P} for complementary descriptions
of the binary hardening and ejection process).

\subsubsection{Eccentric mergers forming during hardening}\label{sec:Eccentric mergers forming during hardening}

We now estimate the probability for a BBH to undergo a GW capture merger with an initial $r_{\rm p} < r_{\rm EM}$ (eccentric GW capture merger),
during the binary-single interactions that harden it from its initial SMA $a_{\rm in}$ to its final ejection value $a_{\rm ej}$.
A probability we refer to as ${P}_{\rm EM}(a_{\rm in}, a_{\rm ej})$.
By using that the differential eccentric merger probability can be written as $dP_{\rm EM}(a) = P_{\rm EM}(a)dN_{\rm bs}$, together
with $da = -a(1-\delta)dN_{\rm bs}$, one finds,
\begin{equation}
{P}_{\rm EM}(a_{\rm in}, a_{\rm ej}) = \frac{1}{1-\delta}\int_{a_{\rm ej}}^{a_{\rm in}} \frac{P_{\rm EM}(a)}{a}da \approx \frac{P_{\rm EM}(a_{\rm ej})}{1-\delta},
\label{eq:P_EM_ain_aej}
\end{equation}
where for the last term we have assumed that $a_{\rm in} \gg a_{\rm ej}$.
As seen, in this limit ${P}_{\rm EM}$ does not depend on $a_{\rm in}$, i.e., our estimate is not strongly dependent on the initial conditions of the
BBH and how it exactly formed. For $\delta  = 7/9$, we therefore conclude that our model, although idealized, seems to robustly predict that the series of
hardening binary-single interactions the BBH must undergo before ejection, leads to a relative increase in the eccentric
GW capture merger probability by a factor of $\approx 9/2$, compared to simply evaluating the probability at $a_{\rm ej}$.

\subsection{Relation to cluster compactness}

The value of ${P}_{\rm EM}(a_{\rm in}, a_{\rm ej})$ depends on $a_{\rm ej}$, which we note in turn depends
on the cluster environment through its escape velocity $v_{\rm esc}$.
By using the relations for $\Delta{E}_{\rm bs}$ presented back in Section \ref{sec:Binary-single hardening process}, and
that $v_{\rm B} \approx v_{\rm esc}$ when $a \approx a_{\rm ej}$, per definition, one finds the following relation,
\begin{equation}
a_{\rm ej} \approx \frac{1}{6} \left(\frac{1}{\delta} - 1\right) \frac{Gm}{v_{\rm esc}^2}.
\end{equation}
The probability ${P}_{\rm EM}$ is therefore $\propto v_{\rm esc}^2$, leading to the general result that the higher $v_{\rm esc}$ is, the higher
${P}_{\rm EM}$ is. Using that the escape velocity relates to the cluster compactness as $v_{\rm esc}^2 \approx GM_{\rm C}/R_{\rm C}$,
where $M_{\rm C}$ and $R_{\rm C}$ denote the characteristic mass and radius of the cluster, respectively, one finds,
\begin{equation}
{P}_{\rm EM}(a_{\rm in}, a_{\rm ej}) \approx \frac{12 \delta N_{\rm IMS}}{(1-\delta)^{2}}\frac{r_{\rm EM}}{m} \times \frac{M_{\rm C}}{R_{\rm C}}.
\label{eq:PEM_MCRC}
\end{equation}
This leads to the important conclusion that the fraction of BBHs that undergoes an eccentric GW capture merger before being ejected from the cluster,
increases linearly with the compactness of the cluster. Measuring the fraction of eccentric to circular merges can therefore be used to
probe the environmental origin of BBH mergers, as described later in Section \ref{sec:Rate of Eccentric Mergers}.

In addition, this further suggests that GW capture mergers could play a significant dynamical role in relative compact clusters, as they are intrinsically formed
inside and bound to the cluster in contrast to the ejected population. If this would lead to a run-away BH build up, or unique
GW observables, is straight forward to study with full $N$-body simulations including PN effects (as with the MC
cluster studies \citep{2016PhRvD..93h4029R, 2017MNRAS.464L..36A}, recent $N$-body studies on BH dynamics in clusters
do not include PN terms \citep{2017MNRAS.469.4665P}). We reserve that for a future study.

\subsection{Three-body vs. two-body mergers}

So far we have only considered the probability for a BBH to undergo a merger inside the cluster
during three-body interactions; however, a non-negligible fraction of the BBHs will undergo
a two-body merger inside the cluster between interactions, or outside the cluster after being ejected.
In the sections below we start by comparing the probability for a BBH to undergo an eccentric merger inside the cluster
doing three-body interactions to the probability that an ejected BBH merger is eccentric. We perform this comparison
to illustrate the importance of the three-body GW capture mergers considered in this work, and thereby the inclusion of PN terms in the $N$-body EOM.
We then estimate the probability for a BBH to undergo an isolated two-body merger between interactions inside the cluster before dynamical ejection is possible.
Finally we list how these different merger types and outcomes scale with the BH mass and host cluster properties.

\subsubsection{Importance of three-body mergers and PN terms}\label{sec:Fraction of eccentric mergers}

Cluster simulations that are based on Newtonian codes \citep[\textit{e.g.},][]{2016PhRvD..93h4029R, 2017MNRAS.464L..36A}, are in principle only
able to probe the population of BBHs that merge outside the cluster after being dynamically ejected; however, the ejected BBHs
are not representative for the population of eccentric BBH mergers \citep[\textit{e.g.},][]{2017ApJ...840L..14S}. Recent Newtonian studies have therefore
vastly underestimated the fraction of BBH mergers that will appear eccentric at the time of observation. To quantify how many more eccentric BBH mergers
that is expected to form when our considered GW capture mergers are taken into account, we first use that the probability
for an ejected BBH to have $r_{\rm p} < r_{\rm EM}$, i.e. to appear with an $e > e_{f}$ at frequency $f$, denoted here by $P_{\rm EM}^{\rm ej,bin}(a_{\rm ej})$,
is simply given by $P_{\rm EM}(a_{\rm ej})/N_{\rm IMS}$. This leads us to the following ratio,
\begin{equation}
\frac{{P}_{\rm EM}(a_{\rm in}, a_{\rm ej})}{{P}_{\rm EM}^{\rm ej,bin}(a_{\rm ej})} = \frac{N_{\rm IMS}}{1-\delta} \approx 100,
\label{eq:fracinc_eccm}
\end{equation}
which states that if one takes into account the eccentric GW capture mergers
then the probability for forming an eccentric BBH merger is about \emph{two orders of magnitude higher} than one finds from only
considering the ejected BBH population. This clearly illustrates the importance of including PN terms.

\subsubsection{Isolated two-body mergers between interactions}\label{sec:Isolated Merger Before Ejection}

Our approach for calculating the total probability for a BBH to undergo an eccentric merger during its hardening binary-single interactions, relies on the assumption
that the probability for it to undergo a merger before ejection is $\ll 1$.  However, in very dense stellar systems $a_{\rm ej}$ will be so low that the
BBH has a non-negligible probability to merge between its binary-single interactions before the ejection limit is reached \citep[\textit{e.g.},][]{2016ApJ...831..187A}.
In the following we estimate the probability for a BBH to merge between interactions integrated from $a_{\rm in}$ to $a_{\rm ej}$,
a probability we denote by $P_{\rm IM}(a_{\rm in}, a_{\rm ej})$, where `IM' is short for `Isolated Merger'.

To estimate $P_{\rm IM}(a_{\rm in}, a_{\rm ej})$ we first need to derive the
probability for that a BBH with SMA $a$ undergoes an isolated merger before its next binary-single interaction, $P_{\rm IM}(a)$.
Using that the GW inspiral life time can be written as $t_{\rm life}(a,e) \approx t_{\rm life}(a)(1-e^{2})^{7/2}$ \citep{Peters:1964bc}, where $t_{\rm life}(a)$ denotes
the `circular life time' for which $e=0$, and assuming the BBH eccentricity distribution follows $P(e) = 2e$ \citep{Heggie:1975uy}, one finds that \citep{2018ApJ...853..140S},
\begin{equation}
{P}_{\rm IM}(a) \approx
\begin{cases}
(t_{\rm bs}(a)/t_{\rm life}(a))^{2/7}, & \ t_{\rm life}(a) > t_{\rm bs}(a)  \\[2ex]
1, & \ t_{\rm life}(a) \leq t_{\rm bs}(a),
\end{cases}
\label{eq:Pt}
\end{equation}
where $t_{\rm bs}(a)$ denotes the the average time between binary-single encounters at SMA $a$.
The time $t_{\rm bs}(a)$ is to leading order inversely proportional to the BH binary-single encounter rate, i.e.,
$t_{\rm bs}(a) \approx 1/\Gamma_{\rm bs} \approx (n_{\rm s}\sigma_{\rm bs}v_{\rm disp})^{-1}$,
where $n_{\rm s}$ is the number density of single BHs, $\sigma_{\rm bs}$ is the binary-single interaction cross section,
and $v_{\rm disp}$ is the local velocity dispersion (See e.g. \citep{2018ApJ...853..140S}). With this expression for ${P}_{\rm IM}(a)$, we can now estimate
$P_{\rm IM}(a_{\rm in}, a_{\rm ej})$ by simply integrating ${P}_{\rm IM}(a)$ over the BBH hardening iteractions, in the same
way as we estimated $P_{\rm EM}(a_{\rm in}, a_{\rm ej})$ in Section \ref{sec:Eccentric mergers forming during hardening}. Following this approach we find,
\begin{equation}
{P}_{\rm IM}(a_{\rm in}, a_{\rm ej}) \approx \frac{1}{1-\delta}\int_{a_{\rm ej}}^{a_{\rm in}} \frac{P_{\rm IM}(a)}{a}da \approx \frac{7}{10}\frac{P_{\rm IM}(a_{\rm ej})}{1-\delta},
\label{eq:P_IM_ain_aej}
\end{equation}
where for the last term we have assumed that $a_{\rm in} \gg a_{\rm ej}$. The normalization of this expression is only approximate, as we assume
that each $P_{\rm IM}(a)$ is uncorrelated and neither the local cluster environment nor the BBH change properties between interactions.
A similar expression was derived in \cite{2016ApJ...831..187A}, but using a slightly different approach. We will evaluate ${P}_{\rm IM}(a_{\rm in}, a_{\rm ej})$ 
for different clusters and BH masses in Section \ref{sec:Rate of Eccentric Mergers}.

\subsection{Black hole merger scaling relations}\label{sec:Black hole merger scaling relations}

We conclude this section by here presenting the relevant scalings of the different BBH merger types described so far
including isolated mergers, GW capture mergers, eccentric mergers, and ejected BBHs that merge within a Hubble time.
As above, we assume that the probability for merger before ejection is $\ll 1$, which allow us to treat the different
outcomes as uncorrelated. Solving for the general case will be the topic of future studies.

First, the probability for a BBH to undergo an isolated merger between interactions
is given by Equation \eqref{eq:Pt} and \eqref{eq:P_IM_ain_aej}, which also can be written as,
\begin{equation}
P_{\rm IM}(a_{\rm in}, a_{\rm ej}) \propto n_{\rm s}^{-2/7}m^{-6/7}v_{\rm esc}^{22/7} \propto (M_{\rm C}/m)^{4/7} v_{\rm esc}^{10/7},
\label{eq:sc_PIM}
\end{equation}
where for the last equality we have assumed that $n_{\rm s} \propto (M_{\rm C}/m)R_{\rm C}^{-3}$. As seen, $P_{\rm IM}(a_{\rm in}, a_{\rm ej})$
increases both with the escape velocity, $v_{\rm esc}$, and with the number of single BHs in the core, $N_{\rm s}$. The rather surprising scaling with $N_{\rm s}$
originates from that if $N_{\rm s}$ increases for a fixed $v_{\rm esc}$, then the core have to expand which leads to a decrease in the density and thereby the
binary-single encounter rate.

The probability for that a BBH undergoes a GW capture merger during a binary-single interaction integrated from
$a_{\rm in}$ to $a_{\rm ej}$, denoted by $P_{\rm cap}(a_{\rm in}, a_{\rm ej})$,
is proportional to Equation \eqref{eq:Pef_a}, but with $r_{\rm cap}$ from Equation \eqref{eq:rcap} instead of $r_{\rm EM}$ \citep[\textit{e.g.},][]{2018ApJ...853..140S}. From this follows the relation,
\begin{equation}
P_{\rm cap}(a_{\rm in}, a_{\rm ej}) \propto v_{\rm esc}^{10/7}.
\end{equation}
As seen, $P_{\rm cap}(a_{\rm in}, a_{\rm ej})$ is surprisingly independent of the BH mass $m$, the probability for a GW capture merger to form
during hardening depends therefore only on the compactness of the cluster. By comparing with $P_{\rm IM}$ from Equation \eqref{eq:sc_PIM},
one finds $P_{\rm cap}/P_{\rm IM} \propto (M_{\rm C}/m)^{-4/7}$, which suggests that the number of binary-single GW capture mergers relative to the number
of two-body isolated mergers scales inversely with the number of BHs in the core.

The probability for a GW capture merger to appear eccentric at the time of observation is given by Equation \eqref{eq:P_EM_ain_aej}, which can be written as,
\begin{equation}
P_{\rm EM}(a_{\rm in}, a_{\rm ej}) \propto m^{-2/3}v_{\rm esc}^{2}.
\end{equation}
As described in Section \ref{sec:Rate of Eccentric Mergers} below, the ratio between $P_{\rm EM}$ and the probability for that an ejected BBH undergoes
a merger within a Hubble time, denoted here by $P_{\rm HM}(a_{\rm ej})$, directly relates to the observable fraction of
eccentric mergers. The probability $P_{\rm HM}(a_{\rm ej})$ is given by Equation \eqref{eq:Pt}, but with the Hubble time $t_{\rm H}$ as the time limit instead of the binary-single
encounter time $t_{\rm bs}$ \citep{2018ApJ...853..140S}. From this follows the relation,
\begin{equation}
P_{\rm HM}(a_{\rm ej}) \propto m^{-2/7} v_{\rm esc}^{16/7}.
\end{equation}
This lead us to the ratio $P_{\rm EM}/P_{\rm HM} \propto m^{-8/21}v_{\rm esc}^{-2/7}$, which suggests that
the largest fraction of eccentric BBH mergers is formed in interactions involving lower mass BHs in clusters with a relative low
velocity dispersion. In the section below we include the correct normalizations, from which we are able to estimate the fraction of
eccentric BBH mergers observable by LIGO.

\section{Rate of Eccentric Mergers}\label{sec:Rate of Eccentric Mergers}

\begin{figure}
\centering
\includegraphics[width=\columnwidth]{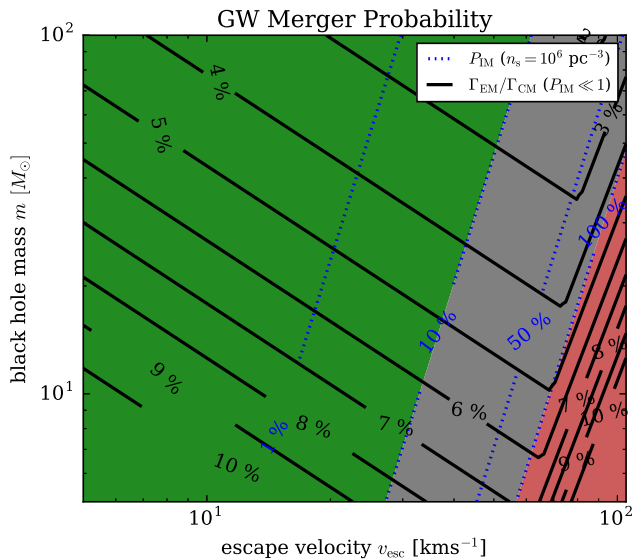}
\caption{The {\it black solid} contours show the ratio ${P}_{\rm EM}(a_{\rm in}, a_{\rm ej})/{P}_{\rm CM}^{<t_{\rm H}}(a_{\rm ej})$,
evaluated for the relevant LIGO values $e_{f} = 0.1$ and $f = 10\ \rm{Hz}$, as a function of the escape velocity of the host
cluster $v_{\rm esc}$ (x-axis), and the BH mass $m$ (y-axis).
As described in Section \ref{sec:Rate of Eccentric Mergers}, this ratio approximately equals the ratio between the present rate of eccentric GW capture mergers
and ejected circular mergers, $R_{\rm E/C} = \Gamma_{\rm EM}/\Gamma_{\rm CM}$, which observationally can be used to
constrain the fraction of all BBH mergers forming in clusters. As seen, the relative rate of BBH mergers with $e \geq 0.1$ at $10$ Hz 
is $\sim 5\%$ for a typical GC, which interestingly suggests that eccentric LIGO sources assembled in clusters will be relative frequent.
The {\it blue dotted} contours show the integrated probability for a given BBH to undergo an isolated merger between encounters during hardening
from SMA $a_{\rm in}$ to $a_{\rm ej}$, ${P}_{\rm IM}(a_{\rm in}, a_{\rm ej}$), derived in Section \ref{sec:Isolated Merger Before Ejection}. For this estimate we
have assumed that $n_{s} = 10^{6}$ pc$^{-3}$ and that $v_{\rm disp} \approx v_{\rm esc}$. The {\it green} region indicates where our estimate of
$R_{\rm E/C}$ is valid (${P}_{\rm IM}(a_{\rm in}, a_{\rm ej}) \ll 0.1$, i.e., merger before ejection is unlikely), the {\it red} region where it breaks down (all BBHs will merge before ejection),
and the {\it grey} region the transition. As seen, our estimate is valid for classical GC systems, but corrections are needed for describing dense nuclear star clusters.}
\label{fig:REC}
\end{figure}

The relevant measure for using eccentric GW mergers to constrain the formation environment of merging BBHs, is not the absolute probability
${P}_{\rm EM}$, but instead the fraction between the rate of eccentric and circular mergers, as this is directly observable,
whereas ${P}_{\rm EM}$ itself is not (${P}_{\rm EM}$ might be indirectly observable if the in-cluster GW capture mergers are able to significantly
alter the cluster dynamics, which could affect the overall BBH merger rate, spin and mass distributions).
For deriving this fraction, we first need to estimate the probability for an ejected BBH to merge within a
Hubble time $t_{\rm H}$, denoted here by ${P}_{\rm CM}^{<t_{\rm H}}$, where
`CM' refers to `Circular Merger' as the ejected population greatly dominates the circular population.
As described in Section \ref{sec:Black hole merger scaling relations} this probability is given by Equation \eqref{eq:Pt}, but with the Hubble
time $t_{\rm H}$ instead of $t_{\rm bs}$ \citep{2018ApJ...853..140S}.
By then assuming that the average rate of binary-single interactions is approximately constant,
one can now approximate the ratio between the present rate of eccentric mergers (forming during binary-single interactions inside the cluster), $\Gamma_{\rm EM}$, and
circular mergers (dominated by the ejected population), $\Gamma_{\rm CM}$, by
\begin{equation}
R_{\rm E/C} = \frac{\Gamma_{\rm EM}}{\Gamma_{\rm CM}} \approx \frac{1}{1-\delta}\frac{{P}_{\rm EM}(a_{\rm ej})}{{P}_{\rm CM}^{<t_{\rm H}}(a_{\rm ej})},
\end{equation}
as further described in \citep{2018ApJ...853..140S}.
The ratio $R_{\rm E/C}$ evaluated for the relevant LIGO values $e_{f} = 0.1$ and $f = 10\ \rm{Hz}$ is shown with black contour lines
in Figure \ref{fig:REC}, as a function of cluster escape velocity $v_{\rm esc}$, and BH mass $m$, where the green colored region roughly indicates where our
estimate for $P_{\rm EM}$ is valid (${P}_{\rm IM}(a_{\rm in}, a_{\rm ej}) < 0.1$ assuming a constant single BH density of $n_{\rm s} = 10^{6}$ pc$^{-3}$).
As seen, our model suggests that $\sim 5\%$ of all observable GW mergers originating from GCs will have an eccentricity $e > 0.1$
when entering the LIGO band for BHs with masses $\lesssim 50\ M_{\odot}$ assembled in a typical GC system.
In more dense environments, such as in galactic nuclei where the escape velocity is significantly higher \citep[\textit{e.g.},][]{2017MNRAS.467.4180S}, our estimate breaks down
as the probability for the interacting BBHs to merge between encounters before ejection is possible is close to unity (red colored region).
Eccentric mergers will still form in such dense environments, but estimating their relative rate requires higher order corrections to our formalism, which will be
the topic of future work. Some work has been done on this limit by \cite{2016ApJ...831..187A}, but without the PN terms we have shown to be crucial in this work.
 
Finally, we do note from Figure \ref{fig:REC} that $R_{\rm E/C}$ does not take unique values across
$v_{\rm esc}$ and $m$, making it difficult to accurately infer the environment based on $R_{\rm E/C}$ alone.
However, it is possible to break this degeneracy by the use of absolute rates, which illustrates both the promising future and
necessity for including GR terms in cluster studies.

\section{Conclusions}
We have in this paper studied the dynamical and GR evolution of BBHs undergoing three-body interactions in dense clusters, from which we find that the rate of eccentric BBH mergers
observable by LIGO (eccentricity $>0.1$ at $10$ Hz) relative to the rate of circular BBH mergers is likely to be $\approx 5\%$ (see Figure \ref{fig:REC}),
for standard GC systems. This eccentric population form through GW captures during resonating
binary-single interactions (Figure \ref{fig:resGWcapex}), and can therefore only be resolved using an $N$-body code that includes the $2.5$ PN term,
which accounts for orbital energy dissipation through the emission of GWs \citep[\textit{e.g.},][]{2014LRR....17....2B}.
This explains why recent Newtonian MC studies \citep[\textit{e.g.},][]{2016PhRvD..93h4029R}, not have been able to resolve this population
(See Section \ref{sec:Fraction of eccentric mergers}). Therefore, despite what have been concluded in the recent literature, our results strongly suggest that
eccentricity can be used to observationally distinguish different BBH merger channels from each other. For example, if no eccentric BBH mergers are observed in the
first, say, 100 LIGO observations, then the field binary channel are likely to be in favor of the GC channel. This greatly motivates recent work on eccentric wave
forms \citep[\textit{e.g.},][]{2017PhRvD..95b4038H, 2017arXiv170510781G, 2018PhRvD..97b4031H}, and might be one of the only reliable tests if the majority of BHs are born with
intrinsic small spin.

Our results further suggest that the rate of GW capture mergers forming during binary-single interactions, eccentric or
not (see Section \ref{sec:GW capture distance}), to the rate of ejected mergers is higher than the
$\sim 2\%$ previously stated \citep{2017ApJ...840L..14S, 2018ApJ...853..140S}, as a newly formed BBH generally undergoes several
interactions before being ejected, and not only one. The relative increase from this hardening process can be found by integrating
the capture probability $P_{\rm cap}$ from $a_{\rm in}$ to $a_{\rm ej}$, similar to the procedure described in Section \ref{sec:Eccentric mergers forming during hardening},
which evaluates to $(7/5)\times(1/(1-\delta)) \approx 6$ for $\delta = 7/9$, suggesting that GW capture mergers forming during binary-single
interactions are likely to constitute $\sim10\%$ of all observable BBH mergers assembled in GCs.
As noted by \cite{2018ApJ...853..140S}, the GW capture mergers will remain bound to their host cluster if the GW kick is low, which
could lead to significant dynamical changes of the cluster at especially early times where the GW capture scenario likely dominates
the BBH merger rate \citep{2018ApJ...853..140S}. These changes could propagate to what we observe today, implying that GW captures
might be indirectly probed even if their current rate is low. This is straight forward to study using a PN $N$-body code and will be the topic of
future studies.

%--------------------------------------------------------------------------
% Acknowledgments%.
%--------------------------------------------------------------------------
\acknowledgments
The author thanks B. Bar-Or, A. Hammers, M. Zaldarriaga, D. Spergel, N. Leigh, M. MacLeod, L. Randell,
E. Ramirez-Ruiz, M. Giersz, A. Askar, B. McKernan, and N. Stone,  for stimulating discussions.
Support for this work was provided by NASA through Einstein Postdoctoral
Fellowship grant number PF4-150127 awarded by the Chandra X-ray Center, which is operated by the
Smithsonian Astrophysical Observatory for NASA under contract NAS8-03060.

\bibliographystyle{h-physrev}
\bibliography{NbodyTides_papers}

\begin{thebibliography}{10}

\bibitem{2016PhRvL.116f1102A}
B.~P. {Abbott} {\em et~al.},
\newblock Physical Review Letters {\bf 116}, 061102 (2016), 1602.03837.

\bibitem{2016PhRvL.116x1103A}
B.~P. {Abbott} {\em et~al.},
\newblock Physical Review Letters {\bf 116}, 241103 (2016), 1606.04855.

\bibitem{2016PhRvX...6d1015A}
B.~P. {Abbott} {\em et~al.},
\newblock Physical Review X {\bf 6}, 041015 (2016), 1606.04856.

\bibitem{2017PhRvL.118v1101A}
B.~P. {Abbott} {\em et~al.},
\newblock Physical Review Letters {\bf 118}, 221101 (2017), 1706.01812.

\bibitem{2017PhRvL.119n1101A}
B.~P. {Abbott} {\em et~al.},
\newblock Physical Review Letters {\bf 119}, 141101 (2017), 1709.09660.

\bibitem{2000ApJ...528L..17P}
S.~F. Portegies~Zwart and S.~L.~W. McMillan,
\newblock \apj {\bf 528}, L17 (2000).

\bibitem{2010MNRAS.402..371B}
S.~{Banerjee}, H.~{Baumgardt}, and P.~{Kroupa},
\newblock \mnras {\bf 402}, 371 (2010), 0910.3954.

\bibitem{2013MNRAS.435.1358T}
A.~{Tanikawa},
\newblock \mnras {\bf 435}, 1358 (2013), 1307.6268.

\bibitem{2014MNRAS.440.2714B}
Y.-B. {Bae}, C.~{Kim}, and H.~M. {Lee},
\newblock \mnras {\bf 440}, 2714 (2014), 1308.1641.

\bibitem{2015PhRvL.115e1101R}
C.~L. {Rodriguez} {\em et~al.},
\newblock Physical Review Letters {\bf 115}, 051101 (2015), 1505.00792.

\bibitem{2016PhRvD..93h4029R}
C.~L. {Rodriguez}, S.~{Chatterjee}, and F.~A. {Rasio},
\newblock \prd {\bf 93}, 084029 (2016), 1602.02444.

\bibitem{2016ApJ...824L...8R}
C.~L. {Rodriguez}, C.-J. {Haster}, S.~{Chatterjee}, V.~{Kalogera}, and F.~A.
  {Rasio},
\newblock \apjl {\bf 824}, L8 (2016), 1604.04254.

\bibitem{2017MNRAS.464L..36A}
A.~{Askar}, M.~{Szkudlarek}, D.~{Gondek-Rosi{\'n}ska}, M.~{Giersz}, and
  T.~{Bulik},
\newblock \mnras {\bf 464}, L36 (2017), 1608.02520.

\bibitem{2017MNRAS.469.4665P}
D.~{Park}, C.~{Kim}, H.~M. {Lee}, Y.-B. {Bae}, and K.~{Belczynski},
\newblock \mnras {\bf 469}, 4665 (2017), 1703.01568.

\bibitem{2012ApJ...759...52D}
M.~{Dominik} {\em et~al.},
\newblock \apj {\bf 759}, 52 (2012), 1202.4901.

\bibitem{2013ApJ...779...72D}
M.~{Dominik} {\em et~al.},
\newblock \apj {\bf 779}, 72 (2013), 1308.1546.

\bibitem{2015ApJ...806..263D}
M.~{Dominik} {\em et~al.},
\newblock \apj {\bf 806}, 263 (2015), 1405.7016.

\bibitem{2016ApJ...819..108B}
K.~{Belczynski} {\em et~al.},
\newblock \apj {\bf 819}, 108 (2016), 1510.04615.

\bibitem{2016Natur.534..512B}
K.~{Belczynski}, D.~E. {Holz}, T.~{Bulik}, and R.~{O'Shaughnessy},
\newblock \nat {\bf 534}, 512 (2016), 1602.04531.

\bibitem{2009MNRAS.395.2127O}
R.~M. {O'Leary}, B.~{Kocsis}, and A.~{Loeb},
\newblock \mnras {\bf 395}, 2127 (2009), 0807.2638.

\bibitem{2015MNRAS.448..754H}
J.~{Hong} and H.~M. {Lee},
\newblock \mnras {\bf 448}, 754 (2015), 1501.02717.

\bibitem{2016ApJ...828...77V}
J.~H. {VanLandingham}, M.~C. {Miller}, D.~P. {Hamilton}, and D.~C.
  {Richardson},
\newblock \apj {\bf 828}, 77 (2016), 1604.04948.

\bibitem{2016ApJ...831..187A}
F.~{Antonini} and F.~A. {Rasio},
\newblock \apj {\bf 831}, 187 (2016), 1606.04889.

\bibitem{2017arXiv170609896H}
B.-M. {Hoang}, S.~{Naoz}, B.~{Kocsis}, F.~A. {Rasio}, and F.~{Dosopoulou},
\newblock ArXiv e-prints  (2017), 1706.09896.

\bibitem{2017ApJ...835..165B}
I.~{Bartos}, B.~{Kocsis}, Z.~{Haiman}, and S.~{M{\'a}rka},
\newblock \apj {\bf 835}, 165 (2017), 1602.03831.

\bibitem{2017MNRAS.464..946S}
N.~C. {Stone}, B.~D. {Metzger}, and Z.~{Haiman},
\newblock \mnras {\bf 464}, 946 (2017), 1602.04226.

\bibitem{2017arXiv170207818M}
B.~{McKernan} {\em et~al.},
\newblock ArXiv e-prints  (2017), 1702.07818.

\bibitem{2016PhRvL.116t1301B}
S.~{Bird} {\em et~al.},
\newblock Physical Review Letters {\bf 116}, 201301 (2016), 1603.00464.

\bibitem{2016PhRvD..94h4013C}
I.~{Cholis} {\em et~al.},
\newblock \prd {\bf 94}, 084013 (2016), 1606.07437.

\bibitem{2016PhRvL.117f1101S}
M.~{Sasaki}, T.~{Suyama}, T.~{Tanaka}, and S.~{Yokoyama},
\newblock Physical Review Letters {\bf 117}, 061101 (2016), 1603.08338.

\bibitem{2016PhRvD..94h3504C}
B.~{Carr}, F.~{K{\"u}hnel}, and M.~{Sandstad},
\newblock \prd {\bf 94}, 083504 (2016), 1607.06077.

\bibitem{2016ApJ...832L...2R}
C.~L. {Rodriguez}, M.~{Zevin}, C.~{Pankow}, V.~{Kalogera}, and F.~A. {Rasio},
\newblock \apjl {\bf 832}, L2 (2016), 1609.05916.

\bibitem{2017ApJ...842L...2C}
X.~{Chen} and P.~{Amaro-Seoane},
\newblock \apjl {\bf 842}, L2 (2017), 1702.08479.

\bibitem{2016PhRvD..94b4012H}
I.~{Harry}, S.~{Privitera}, A.~{Boh{\'e}}, and A.~{Buonanno},
\newblock \prd {\bf 94}, 024012 (2016), 1603.02444.

\bibitem{2017PhRvD..95b4038H}
E.~A. {Huerta} {\em et~al.},
\newblock \prd {\bf 95}, 024038 (2017), 1609.05933.

\bibitem{2017arXiv170510781G}
L.~{Gond{\'a}n}, B.~{Kocsis}, P.~{Raffai}, and Z.~{Frei},
\newblock ArXiv e-prints  (2017), 1705.10781.

\bibitem{2018PhRvD..97b4031H}
E.~A. {Huerta} {\em et~al.},
\newblock \prd {\bf 97}, 024031 (2018), 1711.06276.

\bibitem{2000ApJ...541..319K}
V.~{Kalogera},
\newblock \apj {\bf 541}, 319 (2000), astro-ph/9911417.

\bibitem{2006ApJ...640..156G}
K.~G{\"u}ltekin, M.~C. Miller, and D.~P. Hamilton,
\newblock \apj {\bf 640}, 156 (2006).

\bibitem{2014ApJ...784...71S}
J.~{Samsing}, M.~{MacLeod}, and E.~{Ramirez-Ruiz},
\newblock \apj {\bf 784}, 71 (2014), 1308.2964.

\bibitem{2013ApJ...773..187N}
S.~{Naoz}, B.~{Kocsis}, A.~{Loeb}, and N.~{Yunes},
\newblock \apj {\bf 773}, 187 (2013), 1206.4316.

\bibitem{2017ApJ...846L..11L}
B.~{Liu} and D.~{Lai},
\newblock \apjl {\bf 846}, L11 (2017), 1706.02309.

\bibitem{2017ApJ...840L..14S}
J.~{Samsing} and E.~{Ramirez-Ruiz},
\newblock \apjl {\bf 840}, L14 (2017), 1703.09703.

\bibitem{2016ApJ...816...65A}
F.~{Antonini} {\em et~al.},
\newblock \apj {\bf 816}, 65 (2016), 1509.05080.

\bibitem{Heggie:1975uy}
D.~C. Heggie,
\newblock \mnras {\bf 173}, 729 (1975).

\bibitem{Hut:1983js}
P.~Hut and J.~N. Bahcall,
\newblock \apj {\bf 268}, 319 (1983).

\bibitem{2014LRR....17....2B}
L.~{Blanchet},
\newblock Living Reviews in Relativity {\bf 17} (2014), 1310.1528.

\bibitem{2016ARA&A..54..441N}
S.~{Naoz},
\newblock \araa {\bf 54}, 441 (2016), 1601.07175.

\bibitem{2018ApJ...853..140S}
J.~{Samsing}, M.~{MacLeod}, and E.~{Ramirez-Ruiz},
\newblock \apj {\bf 853}, 140 (2018), 1706.03776.

\bibitem{2017ApJ...846...36S}
J.~{Samsing}, M.~{MacLeod}, and E.~{Ramirez-Ruiz},
\newblock \apj {\bf 846}, 36 (2017), 1609.09114.

\bibitem{2017arXiv170604672S}
J.~{Samsing} and T.~{Ilan},
\newblock ArXiv e-prints  (2017), 1706.04672.

\bibitem{2017arXiv170901660S}
J.~{Samsing} and T.~{Ilan},
\newblock ArXiv e-prints  (2017), 1709.01660.

\bibitem{Wen:2003bu}
L.~Wen,
\newblock \apj {\bf 598}, 419 (2003).

\bibitem{Peters:1964bc}
P.~Peters,
\newblock Phys. Rev. {\bf 136}, B1224 (1964).

\bibitem{Hansen:1972il}
R.~Hansen,
\newblock Phys. Rev. D {\bf 5}, 1021 (1972).

\bibitem{1983AJ.....88.1549H}
P.~{Hut},
\newblock \aj {\bf 88}, 1549 (1983).

\bibitem{2006tbp..book.....V}
M.~{Valtonen} and H.~{Karttunen},
\newblock {\em {The Three-Body Problem}} (, 2006).

\bibitem{1992ApJ...389..527H}
P.~{Hut}, S.~{McMillan}, and R.~W. {Romani},
\newblock \apj {\bf 389}, 527 (1992).

\bibitem{2017MNRAS.467.4180S}
N.~C. {Stone}, A.~H.~W. {K{\"u}pper}, and J.~P. {Ostriker},
\newblock \mnras {\bf 467}, 4180 (2017), 1606.01909.

\end{thebibliography}

\end{document}